\documentclass[10pt]{iopart}
\usepackage{graphicx}
\usepackage{iopams}
\usepackage{dcolumn}
\usepackage{bm}
\usepackage{color}
\usepackage{epstopdf}

\newcommand{\unit}[2]{\ensuremath{#1\,\mathrm{#2}}}

\newcommand{\reffig}[1]{\fref{#1}}

\newcommand{\density}{\ensuremath{\mathcal{N}}}

\newcommand{\nat}{Nature}
\newcommand{\pra}{Phys. Rev. A}

\begin{document}
\hyphenation{Ryd-berg}
\title[Lifetimes of ultralong-range Rydberg molecules]{Lifetimes of ultralong-range Rydberg molecules in vibrational ground and excited state}

\author{Bj{\"o}rn Butscher, Vera Bendkowsky, Johannes Nipper, Jonathan B Balewski, Ludmila Kukota, Robert L{\"o}w and Tilman Pfau}
\address{
5. Physikalisches Institut, Universit\"{a}t Stuttgart, Pfaffenwaldring 57, 70569 Stuttgart, Germany.}
\author{Weibin Li, Thomas Pohl, Jan Michael Rost}
\address{
Max-Planck-Institut f\"{u}r Physik komplexer Systeme,
Noethnitzer Str. 38, 01187 Dresden, Germany.}
\ead{\mailto{b.butscher@physik.uni-stuttgart.de}}
\date{\today}

\begin{abstract}
Since their first experimental observation, ultralong-range Rydberg molecules consisting of a highly excited Rydberg atom and a ground state atom \cite{2000PhRvL..85.2458G,2009Natur.458.1005B} have attracted the interest in the field of ultracold chemistry \cite{2009Natur.458..975G,Hogan20092931}.
Especially the intriguing properties like size, polarizability and type of binding they inherit from the Rydberg atom are of interest.
An open question in the field is the reduced lifetime of the molecules compared to the corresponding atomic Rydberg states \cite{2009Natur.458.1005B}.
In this letter we present an experimental study on the lifetimes of the \mbox{$^3\Sigma (5s-35s)$} molecule in its vibrational ground state and in an excited state.
We show that the lifetimes depends on the density of ground state atoms and that this can be described in the frame of a classical scattering between the molecules and ground state atoms.
We also find that the excited molecular state has an even more reduced lifetime compared to the ground state which can be attributed to an inward penetration of the bound atomic pair due to imperfect quantum reflection that takes place in the special shape of the molecular potential \cite{2010PhRvL.105p3201B}.

\end{abstract}
\submitto{\jpb}
\maketitle
Ultralong-range Rydberg molecules are a unique kind of molecules that are based on the attractive character of the scattering between a polarizable ground state atom and an electron in a highly excited Rydberg orbit of the second atom.
In contrast to the well known binding mechanisms covalent binding, ionic binding and van der Waals binding, the mechanism responsible for the formation of these molecules is typically not present in nature.
Since their prediction in 2000 \cite{2000PhRvL..85.2458G}, evidence for these exotic molecules has been found \cite{2006PhRvL..97w3002G} and only recently they have been photoassociated from a sample of ultra cold rubidium atoms \cite{2009Natur.458.1005B}.

In a non-perturbative calculation triggered by the experimental findings, we revealed that some of the bound states of these molecules are formed by yet another new binding mechanism based on quantum reflection near a deep abyss of the molecular potential located at a shape resonance in the p-wave scattering cross section of rubidium \cite{2010PhRvL.105p3201B}.
The imperfect quantum reflection at the potential drop may reduce the lifetime of the excited states in excess of the already shorter lifetime of the vibrational ground state compared to the bare atomic state.
In this letter we investigate this open question of the reduced lifetimes and the decay mechanisms leading to this reduction.
We determined the lifetime of vibrational ground state and an excited state of the $35s$-molecules for several densities of ground state atoms and explain the observed behavior in the frame of a classical scattering theory.
We also report on the photoassociation of ultralong-range Rydberg molecules with the Rydberg atom in the $43s$-state in its vibrational ground state and excited states as well as the creation of the trimer state, where the Rydberg atom binds two ground state atoms.
With a bond length of $3000 a_0$ they are the largest observed molecules of this type so far.

\begin{figure}[b!]
\begin{center}
\resizebox{0.99\columnwidth}{!}{\includegraphics{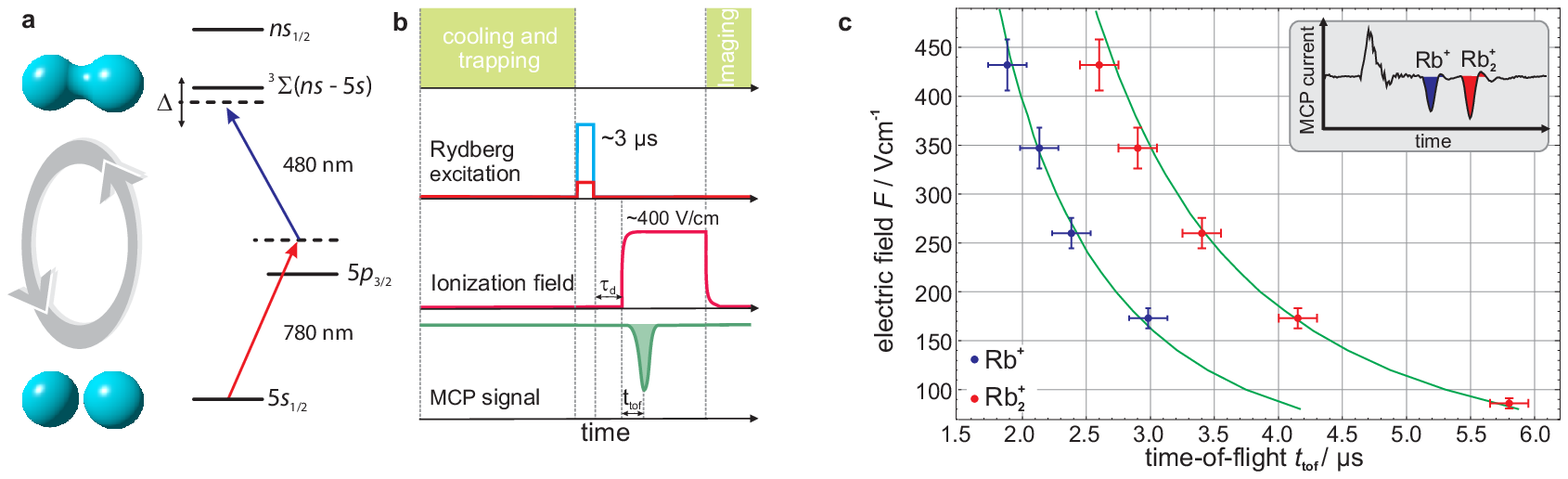}}
\caption{\label{fig1}\textbf{Excitation of ultralong-range Rydberg molecules.}\\
a) Scheme of the photoassociation of ultralong-range Rydberg molecules.
The red laser ($780\,\mathrm{nm}$) is blue detuned from the intermediate $p$-level by ca. $400\,\textrm{MHz}$.
The blue laser frequency ($480\,\mathrm{nm}$) is tuned on two-photon resonance.\\
b) Experimental sequence for the investigation of ultralong-range Rydberg molecules. The cloud of ground state atoms is prepared in a magnetic trap. Then the excitation lasers are switched on for few microseconds. Afterwards, the ionization field is ramped up after a variable delay time $\tau_d$ and the ion signal is detected. In one cloud, up to 400 excitation-detection cycles are possible.\\
c) Calculated and measured time-of-flights for Rb$^+$ (blue) and Rb$_2^+$ (red). Solid lines are calculations for ions with masses of 87 amu and 174 amu, starting from the center of the trap with zero velocity. Measured values are represented by the data points. A typical current signal of the MCP is shown as inset.}
\end{center}
\end{figure}
We create the ultralong-range Rydberg molecules in the \mbox{$^3\Sigma (5s-ns)$} states from a sample of approximately $2\cdot10^6$ $^{87}$Rb atoms at a temperature of $3 \,\mathrm{{\mu}K}$ in the ground state $5s_{1/2}$, \mbox{$F=2; m_F=2$} via a two-photon transition.
The two lasers at $\unit{780}{nm}$ and $\unit{480}{nm}$ are blue detuned by $\approx\unit{400}{MHz}$ from the intermediate $5p_{3/2}$-state.
The laser light is generated by two master-slave diode-laser setups at $\unit{780}{nm}$ and at $\unit{961}{nm}$ being locked to a passively stable cavity.
The $\unit{480}{nm}$ laser light is generated by frequency doubling the $\unit{961}{nm}$ light.
A~schematic diagram of the atomic level structure and the experimental sequence is depicted in \fref{fig1}.
Due to the curvature of the magnetic field and imperfect polarisations, not only the $m_j=+1/2$ magnetic Rydberg substate is excited but also a small fraction of the $m_j=-1/2$ state.
After each laser excitation, we field ionize the created Rydberg atoms and detect the positive ions on a micro-channel plate (MCP).
A subsequent transimpedance amplifier converts the ion current to a voltage signal that we record with a PC-digitizer card.
In a time-of-flight analysis we observe two separate pulses that have similar spectral shape.
We identify these as Rb$^+$ and Rb$_2^+$ ions.
To verify this, we have varied the electric ionization field - and thus the electric field during time-of-flight - and recorded the time delay until we detect the two ion peaks (see \fref{fig1}c).
From the geometry of our chamber and of the field plates that field ionize the Rydberg atoms and guide the ions to the MCP, we also calculated the expected time-of-flight for Rb$^+$ and Rb$_2^+$ using the SIMION software package (green curve in \fref{fig1}c), which is in good agreement with the experimental finding.

\begin{figure}[b!]
\begin{center}
\resizebox{0.99\columnwidth}{!}{\includegraphics{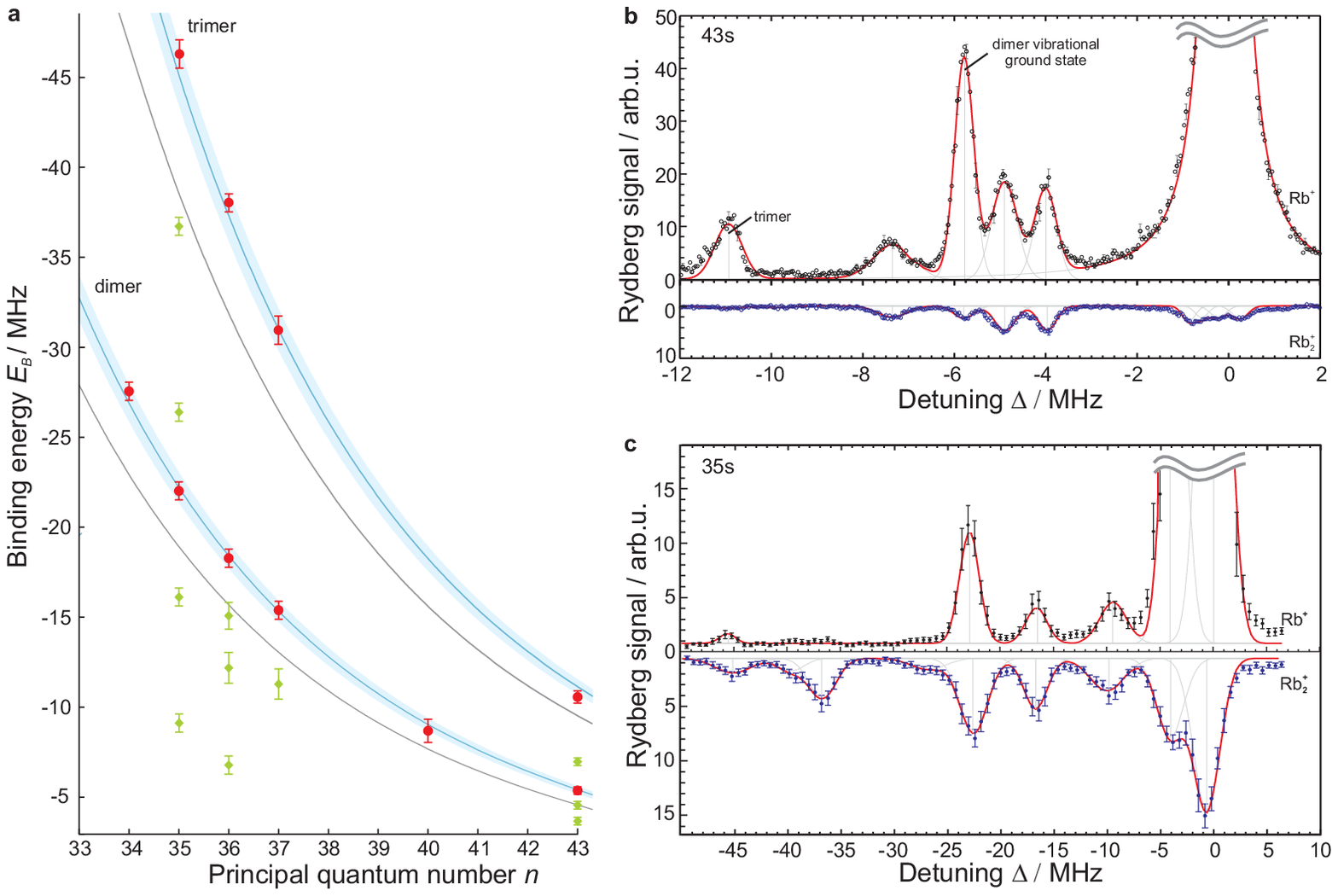}}
\caption{\label{fig2}\textbf{Photoassociation of ultralong-range Rydberg molecules.}\\
a) Measured and calculated binding energies for the ultralong-range Rydberg molecules, $E_B$. The solid blue lines are the calculated binding energies for the vibrational ground states assuming an effective s-wave scattering length $A_\mathrm{eff}=-18.0 a_0$. The shaded area represents the theory for $A_\mathrm{eff}=(-18.0\pm 0.5) a_0$. The grey curves are the calculation for the ab initio value of the s-wave scattering length $A_s=-16.05 a_0$ calculated by Bahrim and Thumm \cite{2000PhRvA..61b2722B}. The red dots mark the line positions that we identify as the vibrational ground state of the dimer or the trimer in the effective s-wave calculation, the green dots are line positions of vibrationally excited states\\
b) Spectrum of the dimer states $^3\Sigma (5s-43s)$ and the corresponding trimer state with magnetic offset field $B_0=7\,\mathrm{G}$. At a detuning of $\Delta=-5.8\,\mathrm{MHz}$ from the atomic resonance we excite the molecular ground state $^3\Sigma (5s-43s) (\nu=0)$.\\
c) Photoassociation spectrum of the dimer $^3\Sigma (5s-35s)$-states and the corresponding trimer state with magnetic offset field $B_0=1\,\mathrm{G}$.\\
Note the different scaling of the frequency axes in b) and c).}
\end{center}
\end{figure}

\begin{figure*}[bth!]
\begin{center}
\resizebox{0.99\columnwidth}{!}{\includegraphics{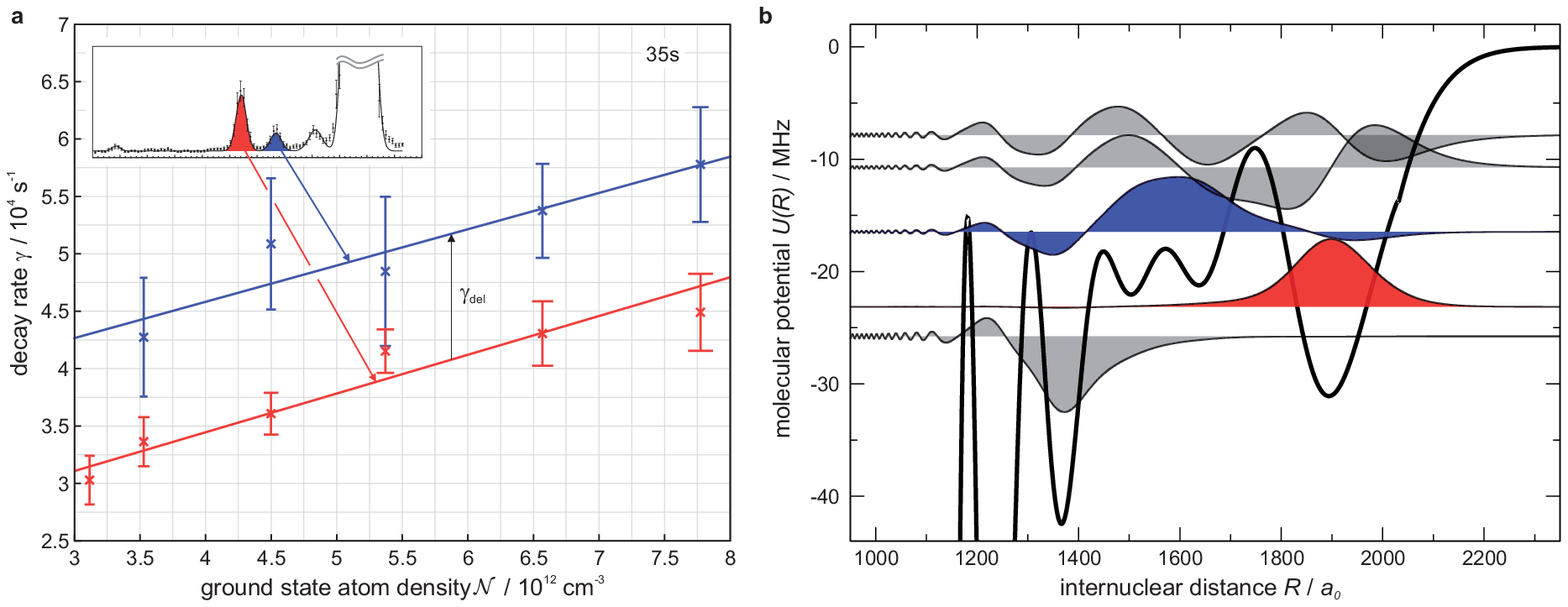}}
\caption{\label{fig3} \textbf{Decay of $^3\Sigma (5s-35s)$ ultralong-range Rydberg molecules.}\\
a) Decay rate of molecular ground state $\gamma_g$ (red) and excited state $\gamma_e$ (blue) vs. density of ground state atoms ${\density}$.
A linear fit of the type $\gamma = \gamma_0+\alpha {\density}$ yields $\gamma_{0,g}=2.1(4) \times 10^4 \mathrm{s}^{-1}, \alpha_g=3(1) \times 10^{-9} \mathrm{cm}^3\mathrm{s}^{-1}$ for the ground state and $\gamma_{0,e}=3.3(12) \times 10^4 \mathrm{s}^{-1}, \alpha_e=3(2) \times 10^{-9} \mathrm{cm}^3\mathrm{s}^{-1}$ for the excited state, approving the linear scaling in $\mathcal{N}$.
The measured decay rate of the Rydberg state $35s$ is $\gamma_\mathrm{atom}=1.6 (2)\times 10^4 \mathrm{s}^{-1}$.
The inset shows the ion signal of the two states in the spectrum (c.f. \fref{fig2}b).\\
b) Molecular potential curve and wave functions for the 35s-state.
The wave function of the vibrational ground state (red) is localized in the outermost potential well (see \fref{fig3}b).
The excited state (blue), however, is spread over several hundred $a_0$ and stabilized by the quantum reflection at the potential drop near $R=1200\,a_0$.
Since the reflection is not perfect, the excited state can reach the inner region with a rate $\gamma_\mathrm{del}$.
}
\end{center}
\end{figure*}

In \fref{fig2}b we show the Rb$^+$- and Rb$_2^+$-signal in the photoassociation spectrum for the $43s$-Rydberg state.
During the excitation, we apply a magnetic offset field of $B_0=7\,\mathrm{G}$.
This shifts the magnetic $m_j=-1/2$ substate of the atom by ca. $20\,\mathrm{MHz}$ to the red which is far enough that it does not affect the observation of the molecular states. 
At a binding energy of $E_B=-5.7\, \mathrm{MHz}$, we find the vibrational ground state of the ultralong-range Rydberg molecules.
Although the potential supports a dimer state which is bound deeper, we denote the state at $E_B=-5.7\, \mathrm{MHz}$ as the vibrational ground state, since we know from previous experiments that the state with the strongest signal is the most localized one in the molecular potential (for analogy at $35s$ see \reffig{fig3}b).
Apart from the vibrational ground state, we also identify two excited molecular states with lower binding energies than the vibrational ground state at $\Delta=-4.9\, \mathrm{MHz}$ and $\Delta=-4.0\, \mathrm{MHz}$ as well as a molecular state with higher binding energy at $\Delta=-7.3\, \mathrm{MHz}$.
At $\Delta=-10.9\, \mathrm{MHz}$, about twice the binding energy $E_B$, we find a trimer state, where one Rydberg atom binds two ground state atoms.

In a previous investigation we found that the binding energies of the excited molecular states can only be calculated correctly in a non-perturbative approach \cite{2010PhRvL.105p3201B}.
We revealed that the p-wave shape resonance leads to an avoided crossing in the molecular potential curve with a deep potential drop near an internuclear separation of $R\approx 1200a_0$.
Surprisingly, the excited states are reflected from the potential drop and are therefore located outside this badland (see \fref{fig3}b).
However, the energies of the vibrational ground states \mbox{$^3\Sigma (5s-ns)(\nu=0)$} are well reproduced in a s-wave theory with an effective scattering length $A_\mathrm{s,eff}=-18.0 a_0$ in the range of $34 \le n \le 40$ \cite{2009Natur.458.1005B}.
In \fref{fig2}a we give an overview of the energies of the bound states that we have identified in the range $34 \le n \le 43$.
To illustrate the validity of the effective s-wave calculation with a modified scattering length up to $n=43$, we also plot the calculated binding energy for an effective scattering length of $A_\mathrm{s,eff}=(-18.0\pm0.5) a_0$ for the dimer and the trimer states.

In the following part of this letter, we will investigate the reduced lifetime of the $^3\Sigma (5s-35s)$ molecular states.
While we found a lifetime of the vibrational ground state of $\tau = 15(3)\mu\mathrm{s}$ in a delayed field ionization experiment \cite{2009Natur.458.1005B}, we found a lifetime of only $\tau = 6.4(9)\mu\mathrm{s}$ when we probed the coherent part of the molecular state \cite{2010NatPh...6..970B}.
However, these results were obtained in different experimental configurations and at different densities of ground state atoms, making it hard to interpret and compare the results.
In order to allow for a systematic investigation of the additional decay channels for ultralong-range Rydberg molecules, we have measured the lifetimes of the molecular ground state and the excited state with binding energy $E_B=-17.4 \,\mathrm{MHz}$ (see \fref{fig3}b) for several values of the density of ground state atoms, $\mathcal{N}$.

To determine the lifetime of the $^3\Sigma (5s-35s)$ molecular states, we use a delayed field ionization technique (see \fref{fig1}b).
We therefore measure the Rydberg population that is still present when the atoms are field ionized at different times $\tau_d$ after turning off the excitation laser.
Since we ramp up the ionization field within $10 \mathrm{ns}$, we do not only ionize the Rydberg state of interest but also states lying close by.
Thus the lifetimes we measure are not reduced by blackbody induced decay.
To eliminate effects resulting from an off-resonant excitation e.g. due to drifts in laser frequency, we simultaneously take two spectra in one atomic sample for each delay time of the field ionization.
For each frequency, we first photoassociate the Rydberg molecules and measure the Rydberg population after delayed field ionization, then we re-excite and measure the Rydberg population with zero delayed field ionization as a reference.
After fitting the two spectra with Gaussian profiles, we obtain the fraction of the Rydberg population remaining at delay time $\tau_d$ as the ratios of the amplitudes of the two fitted curves.
\begin{table}
\caption{\label{tab1}Lifetimes of $^3\Sigma (5s-35s)$ ultralong-range Rydberg molecules in their vibrational ground state ($\tau_g$) and an excited state ($\tau_e$).}
\begin{indented}
\item[]\begin{tabular}{@{}lrrrrrrr}
\br
&\centre{6}{peak density of ground state atoms ${\density} / \mathrm{cm}^{-3}$}\\
& \crule{6}\\
 & $3.1\times 10^{12}$ & $3.5\times 10^{12}$ & $4.5\times 10^{12}$ & $5.3\times 10^{12}$ & $6.6\times 10^{12}$ & $7.8\times 10^{12}$\\
\mr
$\tau_g / \mu\mathrm{s}$  & $33(2)$ & $29(2)$ & $27.7(14)$ & $24.1(11)$ & $23(2)$ & $22(2)$\\
$\tau_e / \mu\mathrm{s}$ & $-$ & $23(3)$ & $21(3)$ & $20(2)$ & $19(2)$ & $17(2)$\\
\\
\br
\end{tabular}
\end{indented}
\end{table}

The measured lifetime of the vibrational ground state and an excited state of the $^3\Sigma (5s-35s)$ molecules for different densities ${\density}$ are given in \tref{tab1}.
For the sake of readability, we will discuss in the following the decay rates $\gamma=1/\tau$.
The decay rates $\gamma_g$ of the molecular ground state (red) and the decay rates of the excited molecular state $\gamma_e$ (blue) are depicted in \fref{fig3}a for different ground state densities $\mathcal{N}$.
The wave functions of the two states in the molecular potentials are shown in \fref{fig3}b.
We find that the decay rates of both states increase as the density of ground state atoms is increased.
We have verified for densities of $5\times 10^{12} \mathrm{cm}^{-3}$ and $15\times 10^{12} \mathrm{cm}^{-3}$ that the decay rate of the bare Rydberg state $35s$ is independent of the density of ground state atoms at a value of $\gamma_\mathrm{atom}=1.6 (2)\times 10^4 \mathrm{s}^{-1}$.
It therefore stands to reason that the decay rate of the molecular states consists of two contributions: the atomic decay rate $\gamma_\mathrm{atom}$, which is present for both molecules and Rydberg atoms and is independent of the density of ground state atoms ${\density}$, and a molecular decay rate $\gamma_\mathrm{mol}$ which is characteristic for the molecular states and depends on ${\density}$.

To model the ${\density}$-dependence of $\gamma_\mathrm{mol}$, we employ classical scattering theory.
We therefore assume that the molecule dissociates when a ground state atom and the molecule collide.
This seems reasonable as an additional ground state atom entering the classical volume $4/3 \pi R^3$ disturbs the Rydberg wavefunction and therefore the molecular potential.
The rate $\gamma_\mathrm{col}$ of such collisions in a gas is a well known problem \cite{maxwell}.
In the case where the molecule can be treated at rest compared to the scattering ground state atoms and the density of molecules is much lower than the density of the ground state atoms, the rate is given by \cite{sommerfeldvorlesungen}
\begin{equation}
\label{gamma_col}
\gamma_\mathrm{col} = {\density} \pi s^2 \overline{v}.
\end{equation}
Here, $\pi s^2$ is the geometric cross section, which in our case is $\pi (r+R)^2\approx \pi R^2$, since the bond length $R$ of the molecule is much larger that the size $r$ of the scattering ground state atom.
The mean relative velocity $\overline{v}$ of atoms with mass $m$ in a thermal sample at temperature $T$ is $\overline{v}=\sqrt{2\cdot 8 k_\mathrm{B}T/\pi m}$ \cite{Niedrig}.
Finally, ${\density}$ is the density of the scattering  particles and it is reasonable to assume that the density of the scatterers is given by the density of ground state atoms.

Applying these assumptions to \eref{gamma_col}, the additional decay rate of the molecular state reads
\begin{equation}
\gamma_\mathrm{col} = R^2 \sqrt{\frac{16 \pi k_\mathrm{B}T}{m}} {\density} = \alpha_\mathrm{col} {\density}, 
\end{equation}
where the parameter $\alpha$ accounts for the mean velocity of the ground state atoms and for the scattering cross section.
For the decay rate of the molecular ground state, we then finally obtain
\begin{equation}
\gamma_g = \gamma_\mathrm{atom}+\alpha_g {\density}.
\end{equation}

To compare our model with the experimental results, we fit a linear curve of the type  $\gamma_g = \gamma_{0,g}+\alpha_g {\density}$ to the data.
For the molecular ground state, we find the best fit for the values $\gamma_{0,g}=2.1(4) \times 10^4 \mathrm{s}^{-1}$ and $\alpha_g=3(1) \times 10^{-9} \mathrm{cm}^3\mathrm{s}^{-1}$ (red curve in \fref{fig3}).
The offset decay rate $\gamma_{0,g}$ is close to the value we expect from our model, the atomic value $\gamma_\mathrm{atom}$.
To calculate the theoretical prediction, $\alpha_\mathrm{col}$, we use the bond length of $R = 1900 a_0$.
For $^{87}$Rb and a temperature of $T=3\mu\mathrm{K}$, we get $\alpha_\mathrm{col}=1.2\times 10^{-9} \mathrm{cm}^3\mathrm{s}^{-1}$, which is in reasonable agreement with the experimental findings.

Until now we have considered the molecular ground state, where the wavefunction is well confined in the outermost potential well.
Fitting a linear curve of the type $\gamma_e = \gamma_{0,e}+\alpha_e {\density}$ for the excited molecular state, we find $\alpha_e=3(2) \times 10^{-9} \mathrm{cm}^3\mathrm{s}^{-1}$, which is the same as for the molecular ground state.
However, the zero-density offset value is $\gamma_{0,e}=3.3(12) \times 10^4 \mathrm{s}^{-1}$.
To understand this shorter lifetime, we have to consider a decay process for the excited state that does not occur for the vibrational ground state.
\begin{table}
\caption{\label{tab2}Lifetimes of bound states with binding energy $E_B$ due to transmission into the inner region of the molecular potential.}
\begin{indented}

\item[]
\begin{tabular}{@{}lrrrrr}
\br
binding energy $E_B$ / MHz & $-7.8$& $-10.7$& $-16.5$& $-23.1$& $-25.9$\\
\mr
Wigner delay time $\tau_\mathrm{del}$ / $\mu$s & $8.5$ & $26.6$ & $40.5$ & $\gg \tau_\mathrm{atom}$ & $14.7$\\
decay rate $\gamma_\mathrm{del}=\tau_\mathrm{del}^{-1}$ / $10^4\mathrm{s}^{-1}$ & $11.8$ & $3.76$ & $2.47$ & $\ll \gamma_\mathrm{atom}$ & $6.80$\\
\br
\end{tabular}
\end{indented}
\end{table}
Figure \ref{fig3}b shows the vibrational wavefunction of the longer lived molecular state, which is well localized in the outermost well of the oscillating potential.
In addition we find several excited states that are delocalized over the outer potential region and bound only by quantum reflection at the deep potential drop arising from the p-wave scattering resonance of the Rydberg-electron ground state atom interaction \cite{2010PhRvL.105p3201B}.
Due to imperfect reflection there is, however, a small probability that these states penetrate to small distances, which ultimately leads to decay of the molecule.
On the other hand, this allows to treat the inner region as an open boundary, such that the excited molecular states can be identified as resonances of an inversed, i.e. inward, scattering process of the two nuclei \cite{2010PhRvL.105p3201B}.
From the asymptotic phase shift $\theta(E)$ one obtains the Wigner delay time
\begin{equation}
\tau_{\rm del}=\frac{\mathrm{d}\theta}{\mathrm{d}E}
\end{equation}
whose peak positions correspond to the location of the quasi-bound molecular states and the corresponding peak amplitude gives the characteristic time of exponential decay to small distances and hence decay of the molecule.
A list of the calculated lifetimes $\tau_{\rm del}= 1/\gamma_{\rm del}$ for $^3\Sigma (5s-35s)$ molecules is given in \tref{tab2}.
For the excited molecular state at $E_B=-16.5\,\mathrm{MHz}$, this yields a decay rate of $\gamma_{\rm del}=2.5\times 10^4\,\mathrm{s}^{-1}$ and, including our measured lifetime of the atomic state, the calculated total decay rate is $\gamma_e=\gamma_\mathrm{atom}+\gamma_\mathrm{del}=4.1\times 10^4\,\mathrm{s}^{-1}$.
Our experimental value $\gamma_{0,e}=3.3(12)\times 10^4\,\mathrm{s}^{-1}$ is slightly lower than the calculated value $\gamma_e$, but still matches within the error bar.
The agreement in the frame of our simple theoretical model is remarkable and is another strong indication that the binding mechanism of the excited state is based on quantum reflection.


In this letter we have presented a systematic study of the lifetime of \mbox{$^3\Sigma (5s-35s)$} ultralong-range Rydberg molecules in their ground state and in an vibrationally excited state.
We could reveal that the lifetime of both states depends on the density of atoms in the ground state $5s_{1/2}$.
The agreement of the experimental results with a classical scattering theory suggests that the reduced lifetime can be explained as a disturbance of the molecular state by the surrounding ground state atoms.
The dependence of this reduced lifetime on temperature $T$ and density $\mathcal{N}$ of the ground state atoms, $\tau_\mathrm{col}\propto T^{1/2}\mathcal{N}^{-1}$, shows again that the ultralong-range Rydberg molecules are observable only in a narrow window of experimental parameters: If the density is too low, no pairs of ground state atoms are close enough to form the molecules and if the density is too high, they decay too fast to be observed at all.

Moreover, we have found that the decay rate of the excited molecular state is systematically higher than that of the molecular ground state. We have given evidence that this probably results from inward penetration of the bound atomic pair due to imperfect quantum reflection, leading to the dissociation of the molecule.
The formation of Rb$_2^+$ in the photoassociation spectra may be a consequence of this process and an investigation of the formation of Rb$_2^+$ might allow a deeper insight into the binding mechanism.

Although the model we have presented has a remarkable agreement with the experimental findings regarding its simplicity, this approach should be treated as a fingerpost towards the understanding of the dynamics of ultralong-range Rydberg molecules.
We expect that a more sophisticated model based on the ideas we have outlined in this letter will lead to a quantitative understanding of the lifetimes of the molecules.
The Ramsey-techniques applied to Rydberg molecules recently \cite{2010NatPh...6..970B} might prove as a helpful tool in the experimental exploration of the dynamics of the molecules.

\ack
This work is funded by the Deutsche Forschungsgemeinschaft (DFG) within the SFB/TRR21 and the project PF~381/4-2.

\end{document}